\documentclass[aps,
reprint,
superscriptaddress,
amsmath,amssymb,
prb,
floatfix,
]{revtex4-2}

\usepackage[utf8]{inputenc}
\usepackage[T1]{fontenc}
\usepackage{graphicx}\usepackage{amsmath}
\usepackage{amssymb}
\usepackage{physics}
\usepackage[colorlinks=true,allcolors=blue]{hyperref}
\usepackage{lipsum}
\usepackage{comment,xcolor}
\usepackage{soul}
\usepackage{float}
\usepackage[normalem]{ulem}
\usepackage{epstopdf}
\usepackage{siunitx}
\usepackage{chemformula}
\usepackage{booktabs,multirow}
\graphicspath{{fig/}} \usepackage{nameref}
\usepackage{dcolumn}

\newcommand{\orcid}[1]{\href{https://orcid.org/#1}{\includegraphics[width=8pt]{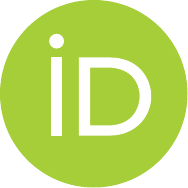}}}

\begin{document}
\title{Machine learning for materials discovery:\\ two-dimensional topological insulators}

\author{Gabriel R. Schleder\orcid{0000-0003-3129-8682}}
\email{gabriel.schleder@ufabc.edu.br}
\affiliation{Federal University of ABC (UFABC), 09210-580, Santo André , São Paulo, Brazil}
\affiliation{Brazilian Nanotechnology National Laboratory (LNNano), CNPEM, 13083-970, Campinas, São Paulo, Brazil}
\affiliation{John A. Paulson School of Engineering and Applied Sciences, Harvard University, Cambridge, Massachusetts 02138, USA}

\author{Bruno Focassio\orcid{0000-0003-4811-7729}}
\email{b.focassio@ufabc.edu.br}
\affiliation{Federal University of ABC (UFABC), 09210-580, Santo André , São Paulo, Brazil}
\affiliation{Brazilian Nanotechnology National Laboratory (LNNano), CNPEM, 13083-970, Campinas, São Paulo, Brazil}

\author{Adalberto Fazzio\orcid{0000-0001-5384-7676}}
\email{adalberto.fazzio@lnnano.cnpem.br}
\affiliation{Federal University of ABC (UFABC), 09210-580, Santo André , São Paulo, Brazil}
\affiliation{Brazilian Nanotechnology National Laboratory (LNNano), CNPEM, 13083-970, Campinas, São Paulo, Brazil}


\begin{abstract}
One of the main goals and challenges of materials discovery is to find the best candidates for each interest property or application. Machine learning rises in this context to efficiently optimize this search, exploring the immense materials space, consisting of simultaneously the atomic, compositional, and structural spaces.
Topological insulators, presenting symmetry-protected metallic edge states, are a promising class of materials for different applications. However, further, development is limited by the scarcity of viable candidates.
Here we present and discuss machine learning--accelerated strategies for searching the materials space for two-dimensional topological materials. We show the importance of detailed investigations of each machine learning component, leading to different results.
Using recently created databases containing thousands of \textit{ab initio} calculations of 2D materials, we train machine learning models capable of determining the electronic topology of materials, with an accuracy of over 90\%. 
We can then generate and screen thousands of novel materials, \textit{efficiently} predicting their topological character without the need for \textit{a priori} structural knowledge.
We discover 56 non-trivial materials, of which 17 novel insulating candidates for further investigation, for which we corroborate their topological properties with density functional theory calculations.
This strategy is 10$\times$ more efficient than the trial-and-error approach while few orders of magnitude faster and is a proof of concept for guiding improved materials discovery search strategies.
\end{abstract} 

\maketitle 

\section{Introduction}

In materials science, two of the main important and interrelated goals are materials discovery and design. Depending on the context, their meaning can vary. We can broadly define \textit{materials discovery} as the process of finding the maximum number of novel materials towards evaluating their different properties for further consideration and optimization for applications. It is, therefore, an expansive exploration task. 
On the contrary, materials design or optimization starts from a predefined goal, and employ corresponding principles and constraints to select promising candidates, investigating how to further improve a property or feature of interest. 
It is, therefore, a converging exploitation task. 
Most discovery tasks are not purely exploratory and can incorporate a character of directed search in its process, a characteristic of materials design. Therefore, materials discovery and design are usually related in practice, requiring a balance between exploration and exploitation \cite{Shahriari2016}.

The dual nature of materials and their consequent properties embodies a direct relationship where the attributes of the material determine its properties, i.e., $Material \longleftrightarrow Properties$. In turn, materials are defined in terms of their constituent \textit{A}toms (elements), \textit{C}omposition (stoichiometry or ratio of elements), and \textit{S}tructure (in different scales, from crystallographic, micro and nanoscale, to atomic short-range \cite{Souza2021}), forming the acronym \textit{ACS} \cite{Zunger2018}.

Both materials discovery and design are then a problem defined in this high-dimensional space of $ACS$: $Target \ properties=f(Materials)=f(ACS)$.
We can therefore associate the process of discovery and design with an optimization problem, each having different goals: \textit{i)} discovery corresponds to a global search or exploration: understand with sufficient accuracy this combinatorial space as a whole, with the least amount of calculations/evaluations performed; and then \textit{ii)} design corresponds to a local search or exploitation: further explore/optimize locally the most promising regions. This problem is closely related to structure prediction for materials discovery and design by sampling potential energy surfaces \cite{Oganov2019}. 
We here mainly focus on the materials discovery step, even though the results of selected candidates can be the start of further local exploitation.

In the traditional approach, materials are evaluated in a trial-and-error Edisonian fashion one at a time, making the search of the materials space both time- and resource- inefficient. 
With the advances in experimental, information, and computational technologies together with software and methodological developments, a step was given in accelerating search campaigns, by the use of automated strategies to test thousands to millions of candidates, usually in a combinatorial way, instead of case-by-case investigations. This so-called high-throughput (HT) approach \cite{Curtarolo2013,Schleder-MLreview, Schleder2020_perspective} thus results in a large landscape of the materials--properties relationship, enabling further investigation of promising systems by screening methods or inverse design strategies \cite{Zunger2018,Sanchez-Lengeling2018}.
High-throughput methods emerge as a powerful and necessary tool in the new paradigm of data-driven materials science, aiding the creation of consistent databases. Recently HT simulations and experiments are becoming more numerous and accurate, satisfying the need for a reliable source of information and aiding the creation of well-grounded databases \cite{Jain2013, aflowlib, Saal2013, Draxl2019}.

Even though the HT approach has drastically improved the time efficacy aspect of the search, it does not tackle resource efficiency, i.e., in the generation of the candidates' catalog, the target property is not necessarily considered, still being a brute-force approach. This is especially true if constructing general-purpose databases unless the entire materials space could be sampled. Complete sampling is impossible due to the combinatorial explosion and the unknown regions of this space, with novel regions and materials being constantly discovered. 
For every property one is interested in, the sampling has to start from zero if it is not yet present in the database. However, even if it is present, 
there is a potential risk of only a small portion of the resulting candidates having the desired property or functionality. Therefore, beyond the screening of fixed databases, the proper discovery of novel materials is always needed.

Consequently, accelerated strategies are needed to sample the materials space thoroughly and efficiently in directed searches of novel materials with promising properties for different applications, using the feedback from past collected data. 
Fortunately, statistical learning has developed solutions for these problems, not being applied in many areas only due to insufficient available data, which is being overcome in recent years. 
Machine learning (ML) uses available data to identify patterns and exploit the insights extracted towards supporting decision-making processes under uncertainty and predicting unknown data \cite{Schleder-MLreview,Butler2018}. 
Machine learning is being increasingly employed in the context of materials and is already established in tasks such as the creation of atomistic force fields \cite{Deringer2019,Zuo2020}, property prediction \cite{Ramprasad2017, Schleder-MLreview}, and, more recently, active learning and closed-loop autonomous experiments \cite{Kusne2020,Attia2020,Burger2020,Montoya2020,Dunn2019}.

\begin{figure}[ht]
    \centering
    \includegraphics[width=\linewidth]{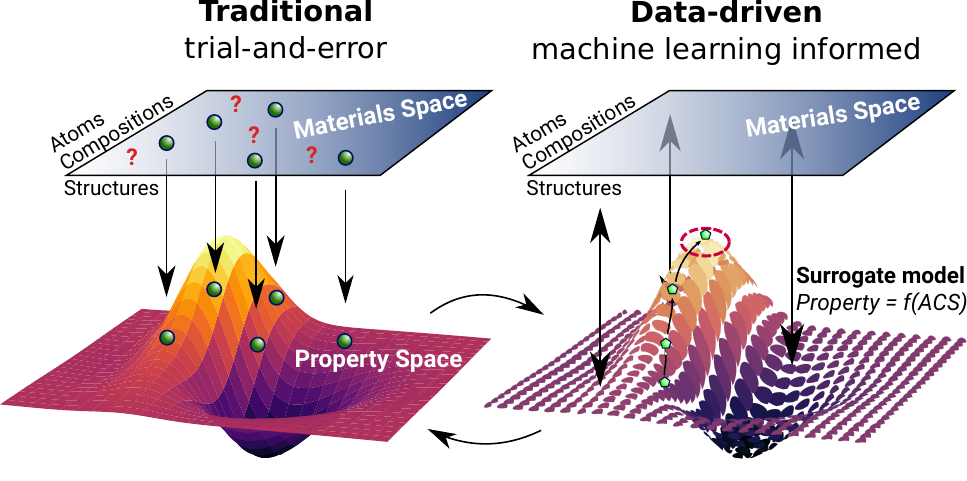}
    \caption{The traditional and inverse data-driven approaches toward sampling a global optimization problem, of which materials discovery is an instance. In the traditional approach, each \textit{a priori} selected material is queried independently to evaluate its corresponding property. The high-throughput approach is a systematic way to accelerate this brute-force search. Conversely, with machine learning, the information already available is used to learn a surrogate function $Property = f(ACS)$, which can then predict the property for any other unknown material. The model is more accurate as more representative information is used to train it.}
    \label{fig:inverse}
\end{figure}

For materials discovery, machine learning uses different methods and algorithms to learn the general function presented earlier, $Target=f(ACS)$, represented in Fig. \ref{fig:inverse}. 
Examples of different used approaches, usually restricted to a specific region of the materials space, are given in  \cite{Saal2020,Schleder-MLreview} and references therein. 
Differently, general learned relations can be extended to other regions of the materials space recovering well-known pressure-induced phase-transition \cite{Ouyang2018} and to more complex compounds \cite{Bartel2019}. 
As we detail in section \ref{sec:results}, in the learning process, the three interrelated ML components are critical in defining the success of this task: input data, the materials representation into numerical features, and the algorithm selection and training \cite{Schleder-MLreview,Schleder2020_perspective,Giustino2020}.
The representation presents one of the most significant challenges because it must translate the $ACS$ into a compact vector while also preserving the significant physical information of the problem. 
For truly novel predictions and materials discovery, a fundamental difference in ML applications is in the $S$ (structural) part of the materials' description. 
No features such as the structure's relaxed geometry and atomic positions are \textit{a priori} known, as obtaining this information would require additional investigations, precisely the results expected to be replaced by the predictive model. Different strategies target the goal of representing materials with atomic-position-independent descriptors \cite{gabriel_stability,Meredig2014,Ward2016,Ward2017,Kim2018,Jain2018,Jacobsen2019}. Finally, there is a fundamental importance in thoroughly evaluating each component of the ML workflow separately \cite{Schleder2020_perspective} and their combination, as it will significantly affect the final results obtained.

\subsection{Two-dimensional topological insulators}
The topological description of solid-state systems, particularly topological insulators (TIs), has been one of the most important developments of modern quantum materials research \cite{Yang2012,Ando2013a,Hasan2010,Moore2010a,Bansil2016,Giustino2020}.
Topological insulators, including quantum spin Hall insulators (QSHIs) \cite{QSH,Kane2005a}, topological crystalline insulators (TCIs) \cite{TCI}, and higher-order topological insulators \cite{Schindlereaat0346, Costa2020}, feature insulating bulk and symmetry-protected metallic boundary states \cite{Hasan2010,RevModPhys.83.1057}. 
These protected states are robust against scattering and local perturbations such as vacancies \cite{Focassio2020, Pezo2020}, impurities \cite{vannucci2020conductance}, and even in the extreme cases of amorphous systems \cite{Agarwala2017,Costa2019,Focassio2020_amorph}, as long as the protective symmetry is preserved. 

These systems are of much interest in practice due to their potential applications in spintronic and spin-orbitronic devices, low energy loss devices, and quantum computing, offering many advantages related to their robustness, such as dissipationless transport and the possibility of operation at ambient temperatures.
Two-dimensional (2D) materials are a natural choice since they offer the additional benefit of miniaturization. The search and discovery of novel 2D TIs has been an increasingly investigated topic. However, suitable candidates must be proposed to accomplish this goal, which must satisfy several different conflicting criteria. Therefore, it is currently a challenge due to this relative topological rarity of about 1\% \cite{marzari_new_TI,Culcer2020}.

The systematic investigation of two-dimensional systems leads to the creation of large 2D materials databases such as MaterialsWeb \cite{Ashton2017}, C2DB \cite{c2db_2018}, Materials Cloud \cite{Mounet2018}, JARVIS-DFT \cite{Choudhary2017}, and 2DMatPedia \cite{Zhou2019}.
%
The emergence of 2D materials databases with various calculated properties enables further materials selection or screening towards desired targets or goals. Some of the previously presented databases include the calculations of topological invariants, allowing for determining the topological character of different materials.
As such, recent works explored each of these databases in search for the calculated topological materials, as in the case of Materials Cloud \cite{marzari_new_TI}, C2DB \cite{thygesen_HT_new_2D_TI}, JARVIS-DFT \cite{Choudhary2020}, and also a comprehensive combination of different databases with individual literature works \cite{Vishwanath2019_2Dprb}.
Until now, each study restricted itself only to the materials already present in the databases, leaving a significant part of the materials' elemental and configurational space yet unexplored.

Recently, the first study of machine learning-enabled 2D materials discovery was published, allowing the prediction of novel materials without \textit{a priori} structural information, using only atomic properties of the materials as features \cite{gabriel_stability}.
In the field of TIs, machine learning models can directly predict the topological classification of a material given its elemental and structural features, which has been performed in specific classes of systems, such as 2D functionalized honeycomb materials \cite{Acosta2018}, tetradymites \cite{Cao2020}, and bulk 3D materials \cite{Regnault2019}, reaching accuracies beyond 90\%. 
Although they can provide insight and generate predictions for specific classes of materials, they do not allow for the prediction of many novel topological insulators beyond the used database, i.e., they do not allow for general topological materials discovery.

In this work, we perform a machine learning-accelerated discovery of novel 2D topological materials for different materials classes.
Using available 2D materials databases, we create predictive models for the topological properties of novel 2D materials. We investigate the interrelated effects of data selection, feature engineering, and algorithm selection on the resulting models, enabling the generalization of the learned property.
With these, we then perform an extensive verification of the topological character of novel systems by both generating 569 new materials combinations and exploring comprehensive 2D materials databases containing thousands of unexplored systems.
We verify the models' predictions with DFT high-throughput calculations, which corroborates the predictions of 56 novel topological materials. 
Of those, 17 are quantum spin Hall insulators (QSHI), while 39 are topological direct-gap metals. We find 3 novel stable QSHIs with electronic band gaps larger than 450 meV---MoTe, AuHg, and WN---suitable for ambient temperature applications. 
The discovery of novel topological materials enables their consideration in screening strategies for future application proposals. Additionally, our workflow is a proof of concept for designing improved search strategies for different target properties. 

\section{Procedures, Results, and Discussion} \label{sec:results}
\begin{figure*}[ht]
    \centering
    \includegraphics[width=\linewidth]{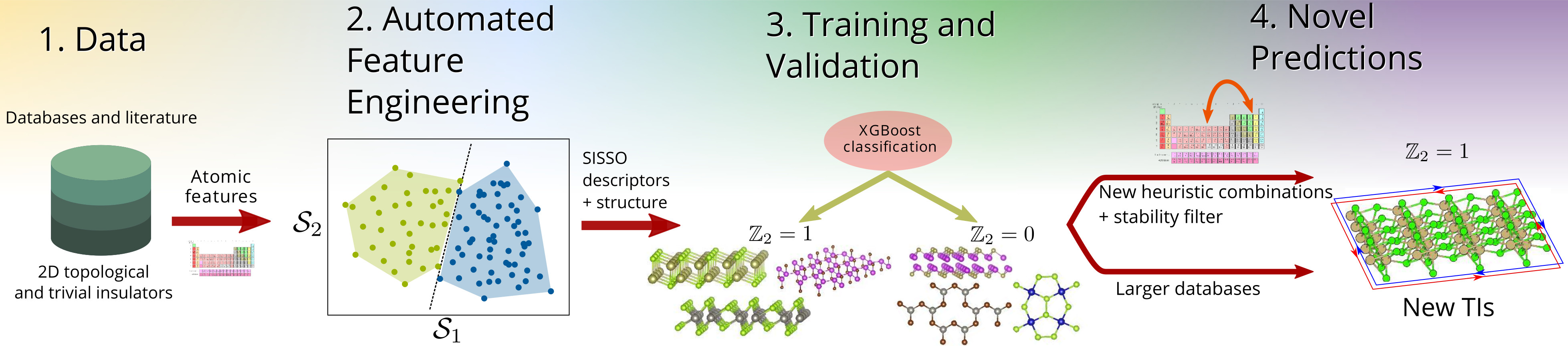}
    \caption{Workflow for learning, prediction, and discovery of new QSHIs. We started by gathering the materials discovered as QSHIs from different databases and references that fulfilled the screening criteria (Fig. \ref{fig:initial_screening}) and used atomic properties as primary features to create our initial dataset. Next, we used the SISSO method to perform an automatic feature engineering of relevant descriptors for the classification problem. Next, the classification task uses the XGBoost method to increase the accuracy and extend prediction capability. Finally, we predict the band topology of novel materials by applying our classifier to new materials generated by heuristic combining elements and more extensive 2D databases that do not have the materials' topological classification.}
    \label{fig:workflow}
\end{figure*}

We can broadly divide the machine learning workflow into four steps \cite{Schleder-MLreview}, which require careful consideration of its details. 
The components are: \textit{i)} problem definition; \textit{ii)} data; \textit{iii)} numerical representation; and \textit{iv)} algorithms, validation, and applications.
Having already established the problem to be tackled, we will explore each of these steps and evaluate the effect of its different choices in the final results obtained.
The general workflow we use is presented in Fig. \ref{fig:workflow}. We train different machine learning models considering changes in each of the variables explored. The models and the corresponding component variables used are given in Table \ref{tab:models}.
Additional models using different combinations of the variables were also evaluated, in most cases leading to decreased performance; a list with the corresponding performance is presented in Table S1 in the Supplementary Material. 
We will now detail each step of the machine learning workflow.

\begin{table*}[]
    \caption{Different machine learning components explored between trained models in order of increasing generality. Prototypes with `+' are grouped according to their similarity. In the magnetism column, `only NM' indicates that only non-magnetic materials are included in the dataset, while `NM gs' indicates that only materials that the non-magnetic configuration is the ground state are included. QSHI are topological insulators with $\mathbb{Z}_2=1$ and DTI are dual topological insulators with $\mathbb{Z}_2=1$ and mirror Chern number $\mathcal{C_M} \neq 0$.}
    \label{tab:models}
    \centering
    \scriptsize
    \begin{tabular}{@{}l p{2cm} p{1.9cm} p{4.9cm} p{1cm} p{1cm} p{1.8cm}  p{3cm}@{}}
    \hline
     &  & \multicolumn{4}{c}{\textbf{Data}} & \multicolumn{2}{c}{\textbf{Representation and algorithm}} \\
     & \# materials & Source   & Prototypes (Fig. \ref{fig:sisso_convex_hull}d) & Mag. & TI type & Atomic features & Structural prototype \\ \midrule
    \textbf{Model A} & 134 NI, 32 TI & Ref. \cite{thygesen_HT_new_2D_TI} & Binaries: C, CH, GaSe, GeSe, ISb, WTe$_2$, BiI$_3$ & Only NM & QSHI & A,B (Table \ref{tab:atomic_properties}) & Multi-task (SISSO) + Simple encoding (XGB) \\
    \textbf{Model B} & 135 NI, 51 TI & Ref. \cite{thygesen_HT_new_2D_TI,marzari_new_TI,Vishwanath2019_2Dprb} & Binaries: C, CH($+$ISb), CH$^*$, GaSe, GeSe, WTe$_2$, GaS, MnS$_2$, ZrTe$_5$ & NM gs & QSHI, DTI & Statistical (Tables \ref{tab:atomic_properties}, \ref{tab:operations}) & Multi-task (SISSO) + Simple encoding (XGB) \\
    \textbf{Model C} & 308 NI, 58 TI & Ref. \cite{thygesen_HT_new_2D_TI,marzari_new_TI,Vishwanath2019_2Dprb} & Binaries+Ternaries: BN($+$C+GeSe), CH($+$ISb), CH$^*$, GaSe, WTe$_2$(+Janus), GaS, MnS$_2$, ZrTe$_5$, MoS$_2$, FeOCl & NM gs & QSHI, DTI & Statistical (Tables \ref{tab:atomic_properties}, \ref{tab:operations}) & Multi-task (SISSO) + Simple encoding (XGB) \\
    \bottomrule
    \end{tabular}
    \end{table*}

\subsection{Materials data screening and automated feature engineering}

\begin{figure*}[!htb]
    \centering
    \includegraphics[width=\linewidth]{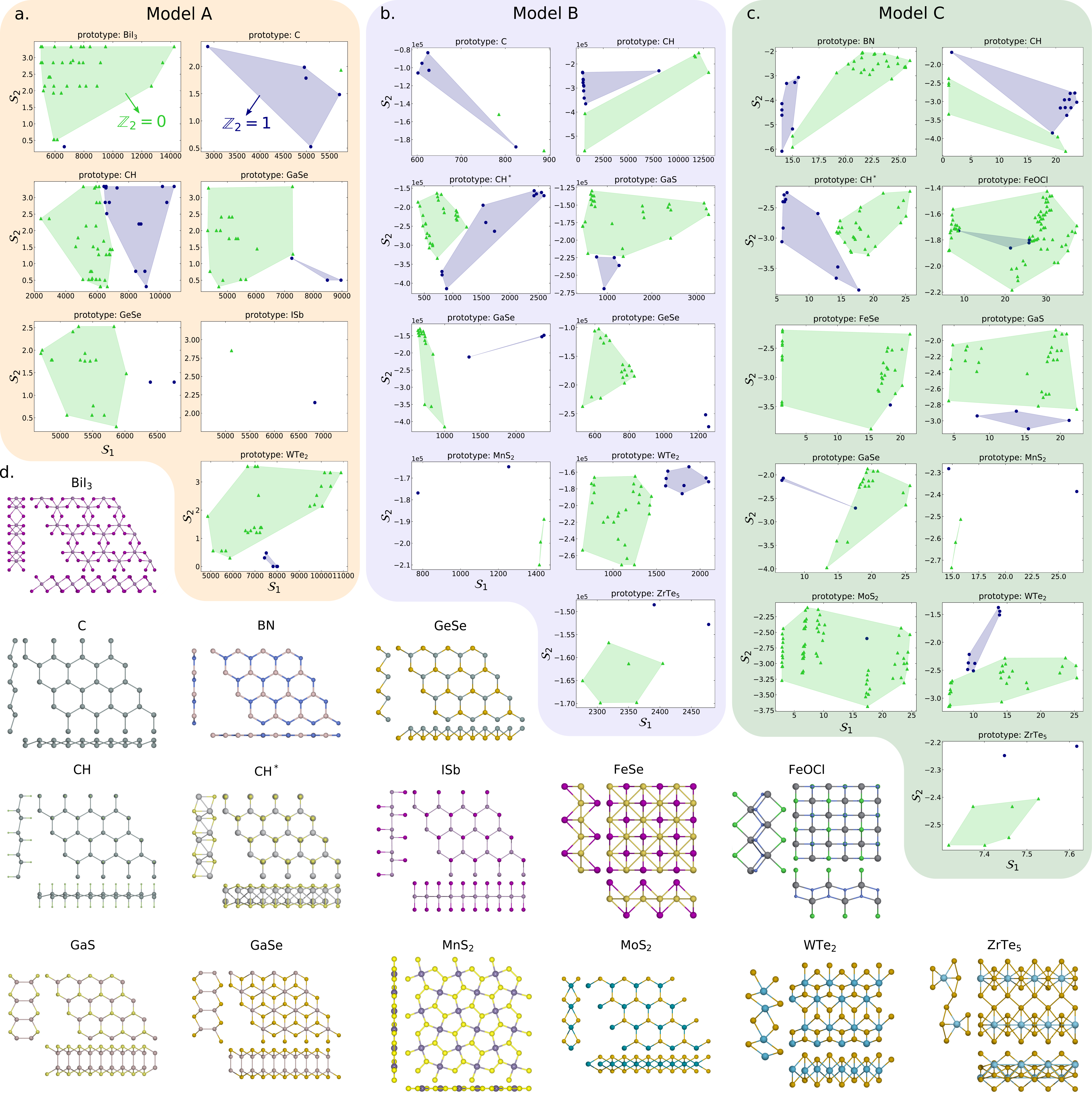}
    \caption{SISSO topological classification maps for the different structural prototypes included in models (a-c) A, B, and C (orange, blue, and green, respectively). The classification convex hull (colored area, trivial and topological materials are represented as light green and dark blue) is obtained with a single pair of multi-task 2D descriptors for each model. The $\mathcal{S}_1$ and  $\mathcal{S}_2$ descriptors obtained from SISSO-MT classification for each model are given in Eqs. (\ref{eq:sisso_desc1})-(\ref{eq:sisso_desc6}).
    (d) Representative materials for each structural prototype considered. Prototypes can be grouped (see Table \ref{tab:models}) according to similarity, such as CH with ISb, or C, BN, and GeSe, representing graphene family (including buckled structures), with a binary basis, and buckled binaries, respectively. }
    \label{fig:sisso_convex_hull} \label{fig:prototypes}
\end{figure*}

The first component for creating the classification model is the input dataset containing examples of both TIs and normal trivial insulators (NIs). Our TI examples are taken from the systematic and openly available data of refs. \cite{thygesen_HT_new_2D_TI,marzari_new_TI,Vishwanath2019_2Dprb}. For the NIs, we use trivial materials from the same structural prototypes containing TIs, taken from the C2DB \cite{c2db_2018} and 2DMatPedia \cite{Zhou2019} databases.
With these, we have a list of materials with their topological classification. 
Each material is classified into a structural prototype. For instance, all \ch{BiX} (X $=$ F, I, Cl, Br) materials present a honeycomb lattice similar to functionalized graphene. Thus, they belong to the structural prototype named CH, which is represented by graphene functionalized with hydrogen atoms, see Fig. \ref{fig:prototypes}d.
Other example is the \ch{WTe2} prototype, which represents the transition metal dichalcogenides (TMDC) 1T$^\prime$ phase, such as \ch{WTe2} and \ch{WSe2}.
These prototypes are attributed to each material in the database (C2DB version 2019.8) creation stage so that this classification can change after DFT geometry optimization. Nevertheless, it is valid a priori information regarding the structure of each material without actually needing the atomic positions, neither the lattice parameters nor layer thickness, which requires DFT calculations.

\begin{figure}[htb]
    \centering
    \includegraphics[width=0.9\linewidth]{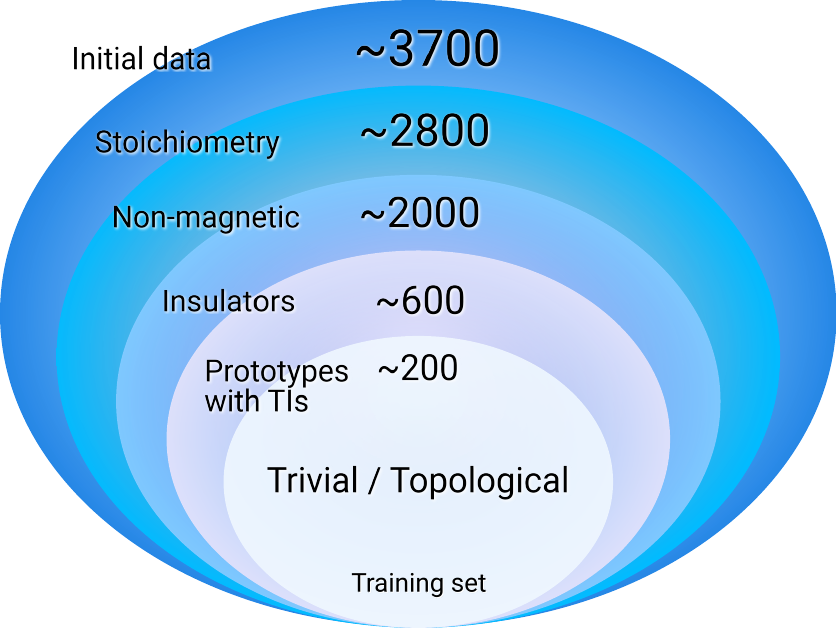}
    \caption{Screening procedure applied to build the materials datasets for the supervised learning task. Starting from references and databases, we screened the binary and ternaries two-dimensional materials that are non-magnetic (NM) insulators from structural prototypes containing at least one instance reported as a TI.}
    \label{fig:initial_screening}
\end{figure}

Although we want our model to be as general as possible, the complexity of the model may be prohibitive for the small database size for this property. Thus, in our first model, hereafter named Model A (Table \ref{tab:models}), we start with a narrower scope by only focusing on materials with hexagonal and TMDC-1T$^\prime$ structural prototypes. Our material list is also screened for binary non-magnetic (NM) insulators whose structural prototype has at least two (except for the prototypes BiI$_3$ and ISb) topological candidates. Therefore we consider the prototypes  \ch{BiI3}, \ch{C}, \ch{CH}, \ch{GaSe}, \ch{GeSe}, \ch{ISb} and \ch{WTe2}. The general screening process, which is also valid for Models B and C, is represented in Fig. \ref{fig:initial_screening}.



The next ingredient is data representation. As motivated in the introduction, we choose to use only atomic properties; these are presented in Table \ref{tab:atomic_properties}. To describe characteristics of the electronic structure usually associated with non-trivial band topology, such as the band inversion, we include the highest occupied and lowest unoccupied Kohn-Sham (KS) eigenvalues with and without SOC, and the total energy difference with and without including SOC.
All quantities are computed using standard DFT calculations detailed in the Methods section.
The other features are experimental and retrieved using the data available in ref. \cite{mendeleev2014}. 

\begin{figure}
    \centering
    \includegraphics[width=\linewidth]{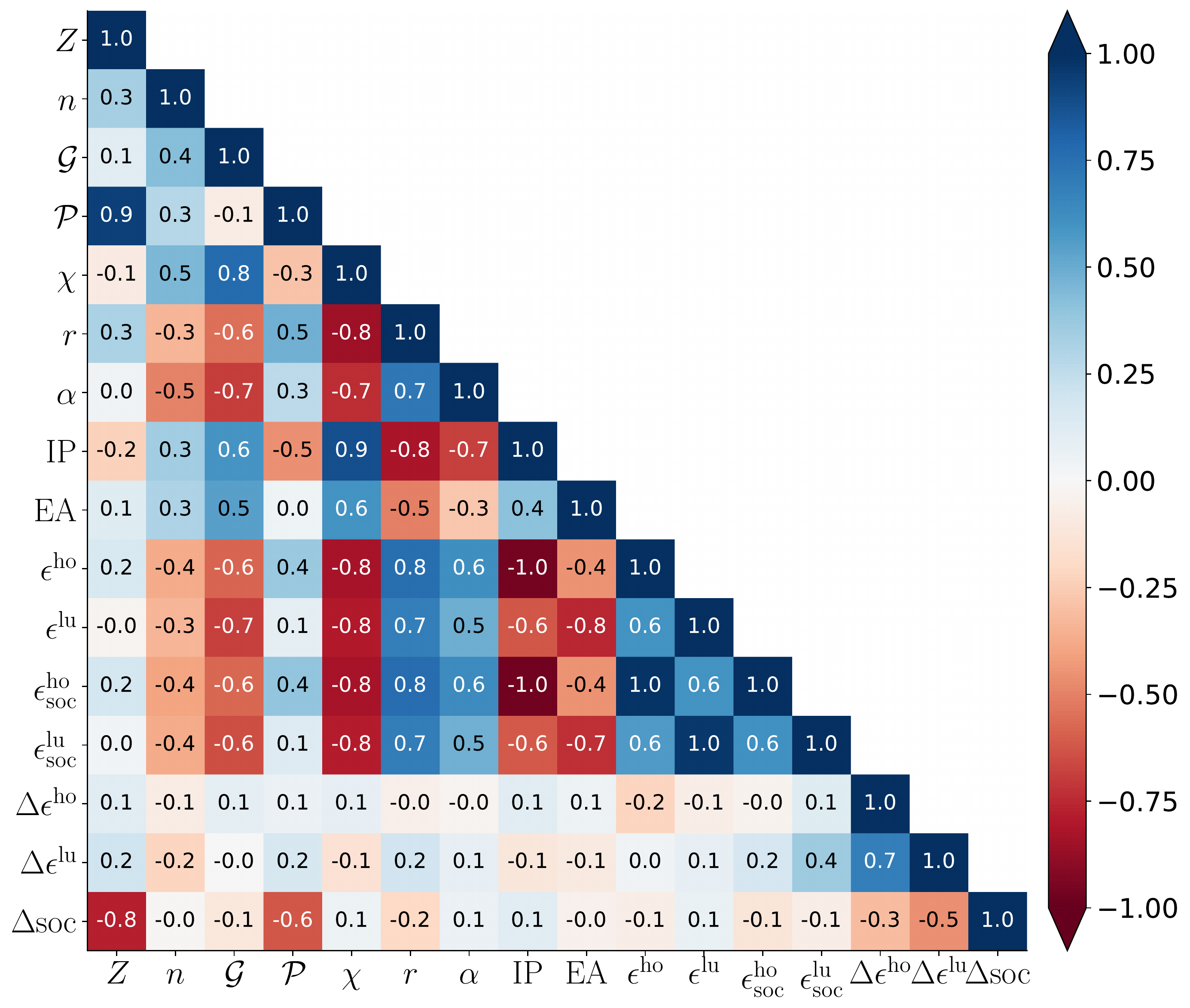}
    \caption{Correlation matrix for the input atomic properties used to construct the primary and subsequent feature spaces. The Pearson's correlation is calculated between atomic features for all elements on the periodic tables up to atomic number 84, except for lanthanides and noble gases.}
    \label{fig:feat_corr}
\end{figure}

The atomic features correlation matrix is presented in Fig. \ref{fig:feat_corr}. The Pearsons'  correlations in Fig. \ref{fig:feat_corr} are computed using the properties of atoms in the periodic table up to atomic number 84, excluding lanthanides and noble gases. Most features have small correlations and provide valuable information for representing different systems. Even correlated features can provide useful information due to nonlinear feature interactions. The correlations between $\epsilon^{\rm \{ho,lu\}}$ and their counterparts with SOC, and with ${\rm EA}$ and ${\rm IP}$ are expected. Also, Fig. \ref{fig:feat_corr} reveals known trends within the periodic table. Importantly, we note correlations between $\Delta{\rm soc}$ and the atomic number ($Z$), which is expected but also important for detecting topological phases. There are small correlations between $\epsilon^{\rm \{ho,lu\}}$ and other properties. However, we expect that using automated feature engineering tools, the combination of $\epsilon^{\rm \{ho,lu\}}_{\rm soc}$ for different atoms may reveal trends for detecting topological phases.

Using the atomic features in Table \ref{tab:atomic_properties}, for model A the data is represented by 34 features, 17 for atom A and 17 for atom B on structural prototypes with formula \ch{A_xB_y}. We use statistical features for models B and C, shown in Table \ref{tab:operations}, that can represent systems of different numbers of elements (unaries, binaries, and ternaries in our case) and compositions \cite{Ward2016}.

%
\begin{table}[!htb]
    \caption{Atomic properties used as input features and their description.}
    \label{tab:atomic_properties}
    \centering
    \begin{tabular}{ll}
    \hline
        \multicolumn{1}{c}{Feature} & \multicolumn{1}{c}{Description} \\ \hline
        $f$ & Element fraction in unit cell \\
        $\chi$ & Pauling electronegativity \\
        $\alpha$ & Atomic polarizability \\
         $r$ & Atom radius \\
         $Z$ & Atomic number \\
        $n$ & Valence \\
        $\mathcal{G}$ & Periodic group \\
        $\mathcal{P}$ & Periodic row$^a$ \\
        EA & Electron affinity \\
        IP & Ionization potential \\
        $\epsilon^{\rm ho}$, $\epsilon^{\rm ho}_{\rm soc}$ & Highest occupied KS eigenvalue \\
          &  without and with SOC \\
        $\epsilon^{\rm lu}$, $\epsilon^{\rm lu}_{\rm soc}$ & Lowest unoccupied KS eigenvalue \\
          &  without and with SOC  \\
        $\Delta \epsilon^{\rm ho}$ & Difference between $\epsilon^{\rm ho}$ and $\epsilon^{\rm ho}_{\rm soc}$ \\
        $\Delta \epsilon^{\rm lu}$ & Difference between $\epsilon^{\rm lu}$ and $\epsilon^{\rm lu}_{\rm soc}$ \\
         $\Delta \rm soc$ & Total energy difference between \\
          & spin-polarized and SOC calculation \\ \hline
    \multicolumn{2}{l}{$^a$ only included in model A} \\
    \end{tabular}
\end{table}

\begin{table}[h!]
    \centering
    \caption{Primary feature space $\Phi_0$ construction using statistical functions for each of the $\gamma$ properties on Table \ref{tab:atomic_properties}, used in Models B and C.}
    \begin{tabular}{cp{6cm}}
    \hline
    \multicolumn{1}{c}{Feature} & \multicolumn{1}{c}{Description} \\ \hline
    $\bar{\gamma}$ & average value $\bar{\gamma}=\sum^{n_{s}}_{i=1}\gamma_{i}/n_{s}$ \\
    $\tilde{\gamma}$ & average weighted by the number of each atom type $\tilde{\gamma}=\sum^{n_{s}}_{i=1}\gamma_{i}n_{i}/N$ \\
    $\gamma_{M}$ & maximum value $\gamma_{M}=\textrm{Max}(\gamma_{i})$ \\
    $\gamma_{m}$ & minimum value $\gamma_{m}=\textrm{Min}(\gamma_{i})$ \\
    $\bar{\gamma}_{\sigma}$ & standard deviation with respect to the average $\bar{\gamma}_{\sigma}=\sqrt{\sum^{n_{s}}_{i=1}(\bar{\gamma}-\gamma_{i})^{2}/n_{s}}$ \\ 
    $\tilde{\gamma}_{\sigma}$ & standard deviation with respect to the weighted average $\tilde{\gamma}_{\sigma}=\sqrt{\sum^{n_{s}}_{i=1}(\tilde{\gamma}-\gamma_{i})^{2}/n_{s}}$ \\
    $[\gamma]$ & range $[\gamma] = \gamma_M - \gamma_m $ \\ \hline
    \end{tabular}
    \label{tab:operations}
\end{table}

Using the atomic features in Tables \ref{tab:atomic_properties} and \ref{tab:operations} and the dataset with screened training examples of TIs and NIs, the next step is to select a learning algorithm to train the classification model. 
We proceed in two stages. First, to perform an automated feature engineering from the simple elemental features selected \cite{gabriel_stability},
we use the multi-task sure independence screening and sparsifying operator (SISSO) method \cite{Ouyang2018,Ouyang2019}. The SISSO method is capable of selecting a multi-dimensional descriptor within an immense set of generated choices, providing a simple feature-based expression to be used for classification, also allowing for physical interpretation \cite{Ghiringhelli2021}.

The SISSO method starts with our primary feature space $\Phi_0$, which contains our input features given in Table \ref{tab:atomic_properties}. Then it constructs a larger feature space $\Phi_1$ by combining the features in $\Phi_0$ using a set of operators $\{I,+,-,\times,\divisionsymbol,\exp,\log,|\;|,\sqrt{\;},^{-1},^2,^3,^6,\cos,\sin\}$ and eliminating the repeated features, note that $\Phi_1$ contains $\Phi_0$. This is sequentially repeated to generate larger feature spaces until reaching the feature space $\Phi_n$, with $n$ set by the user. The method can easily achieve feature spaces with more than \num{e11} features. Then it selects a subspace of descriptors with the largest correlation with the target property by the sure independence screening method (SIS). The optimization to find the best descriptor is then performed using the sparsifying operator (SO). For classification, given that each class forms a convex hull, the optimization of coefficients minimizes the overlap between classes. 
Also, the SISSO method can perform multi-task (MT) learning. In our problem, we want to find a descriptor that translates the atomic features into topological classification. 
The physical understanding of the problem is that structural features are fundamental to band topology but are difficult to be expressed using simple material or atomic properties. 
Nonetheless, we can group the example materials into structural prototypes (groups with a similar structure) and take each prototype as a different task in MT learning. The algorithm finds the best descriptor for all tasks simultaneously, while the minimization of overlap is performed within each task. 

The top-ranked 2D descriptor pair found for Models A, B, and C are expressed in Eqs. \eqref{eq:sisso_desc1} to \eqref{eq:sisso_desc6}, from which we obtain the map for each structural prototypes displayed in Fig. \ref{fig:sisso_convex_hull}. 
This highlights the strength of the SISSO method, finding descriptors in terms of feature combinations, and only using these expressions in most cases we are able to separate both classes, as can be seen in Fig. \ref{fig:sisso_convex_hull}.

\begin{align}
  \mathcal{S}_1^A &= (\text{IP}_A\times\text{IP}_B)\times(\alpha_A + \alpha_B) \label{eq:sisso_desc1} \\
  \mathcal{S}_2^A &= |(\Delta {\rm soc}_A - \epsilon^{\rm lu}_{\rm soc\;A})-(\text{EA}_A \times \chi_A)| \label{eq:sisso_desc2} \\
  \mathcal{S}_1^B &= {\rm IP}_M \times \tilde{r} \times {\rm IP}_m / \mathcal{G}_m \label{eq:sisso_desc3} \\
  \mathcal{S}_2^B &= \mathcal{G}_M^3 \times {\rm IP}_m \times \epsilon_{{\rm soc},m}^{\rm ho} \label{eq:sisso_desc4} \\
  \mathcal{S}_1^C &= \mathcal{G}_m/\bar{f}\times \tilde{\epsilon}^{\rm lu}_{\rm soc}/\epsilon^{\rm lu}_{{\rm soc},m} \label{eq:sisso_desc5} \\
  \mathcal{S}_2^C &= \epsilon^{\rm ho}_m \times \bar{f}+\tilde{{\rm IP}}_\sigma/\chi_M \label{eq:sisso_desc6}
\end{align}
%
%

Equation \eqref{eq:sisso_desc1} highlights that the polarizability ($\alpha$) and ionization potential ($\rm IP$) of A and B atoms are important for defining topological materials. The polarizability increases as the volume occupied by the electrons increases, with larger atoms having more loosely held valence electrons, therefore increasing with the atomic number inside the same column of the periodic table and decreasing from left to right inside the same row of the periodic table due to increase in electronegativity ($\chi$) \cite{Tantardini2021}. The trends for the ionization potential are the opposite. It increases from left to right within a row of the periodic table and decreases with the column in opposition to electronegativity. As seen in Fig. \ref{fig:sisso_convex_hull} for the CH prototype, $\mathcal{S}_1^A$ can partially classify the topological materials; higher values of $\mathcal{S}_1^A$ are non-trivial. This is a reflection of AB materials where B are halides and chalcogenides with medium polarizability and higher ionization potential bonded with A atoms with smaller ionization potential and higher polarizability, especially with A atoms from the \textit{d}-block bonded with chalcogenides, such as ${\rm Ti}_2{\rm X}_2$ (X = S, Cl) and $\rm Zr_2X_2$ (X = S, Se, Cl, Br), also noticing that these are not perfect CH prototypes and in model B falls into the CH$^*$ prototype. This also highlights the importance of revising the prototype classification, which we will discuss later. On the other hand, the descriptor $\mathcal{S}_2^A$ describes two scenarios. First, when the electronegativity and electron affinity ($\rm EA$) are important and second when the spin-orbit coupling energy difference ($\Delta{\rm soc}$) in A atoms dominates.

For the $\rm WTe_2$ prototype, we focus on the region that differentiates between topological and trivial materials within the convex hull map of Fig. \ref{fig:sisso_convex_hull}. When the electronegativity of A atoms is high, the descriptor is balanced by $\Delta{\rm soc}$ and the materials occupy the lower part of the map. In contrast, when the A atoms have high $\Delta{\rm soc}$ and medium electronegativity and electron affinity, $\mathcal{S}_2^A$ is increased, demanding also the $\mathcal{S}_1^A$ for classification, thus favoring higher polarization to differentiate non-trivial from trivial materials. This is evident when considering materials such as $\rm MoX_2$ (X = S, Se, Te) and $WX_2$ (X = Se, Te) in the $\rm WTe_2$ prototype, where the $\mathcal{S}_2^A$ is sufficient for classifying the $\rm MoX_2$ materials and together with $\mathcal{S}_1^A$ can also correctly classify the $\rm WX_2$ materials.

For model B, the prototype revision is critical. Together with the addition of new example materials, we observe two trends contrasted with the model A larger CH prototype. In FIg. \ref{fig:sisso_convex_hull}, there are topological materials classified by a linear combination of higher $\mathcal{S}_2^B$ values and small $\mathcal{S}_1^B$ values, within the more conventional CH prototype and topological materials classified by higher $\mathcal{S}_1^B$ values, entering the CH$^*$ prototype. Most materials within the CH prototype are now composed of heavier \textit{p}-block elements functionalized with chalcogenides and halides. In contrast, the CH$^*$ prototype is mainly composed of \textit{d}-block elements functionalized with chalcogenides. Therefore, smaller $\mathcal{S}_1^B$ values translate the difference in ionization potential, around $3\;\rm eV$, balanced by the smallest value of the periodic table column, which is a higher value when considering topological materials within the CH prototype, consider $\rm Bi_2X_2$ (X = F, Cl, Br, I). This balance is not present in the CH$^*$ prototype because of \textit{d}-block atoms, with comparable ionization potential with the previous case but with elements residing in columns at the leftmost part of the periodic table. Higher $\mathcal{S}_2^B$ values that help to classify the CH prototype are due to elements in the rightmost column of the periodic table, mainly halides, taking the third power for increasing the difference compared to CH$^*$ prototype.

Notably, the $\mathcal{S}_2^B$ descriptor is also capable of correctly classifying GaS materials. Some examples inside the GaS prototype reveal that topological materials have A atoms with the smallest value of the highest occupied KS eigenvalue with SOC ($\epsilon_{\rm soc}^{\rm ho}$). In contrast, trivial materials have $\epsilon_{{\rm soc},\,m}^{\rm ho}$ values of the B atoms, with almost $\rm 1\;eV$ separating $\epsilon_{\rm soc}^{\rm ho}$ values of A and B atoms. In the $\rm WTe_2$ prototype, $\mathcal{S}_1^B$ is the primary descriptor for classification as a result of most topological materials with \textit{d}-block elements located at the leftmost part of the periodic table, such as W, Mo, and Ta, increasing $\mathcal{S}_1^B$. In contrast, trivial materials have elements at the rightmost part of the transition metal block, decreasing $\mathcal{S}_1^B$.

With model C, focusing on binary prototypes, we observe that for both BN and CH$^*$ prototypes, the $\mathcal{S}_1^C$ descriptor is fundamental, decreasing $\mathcal{S}_1^C$ results in topological materials with elements located at the leftmost portion of the periodic table, also reducing the smallest $\epsilon_{\rm soc}^{\rm ho}$ value, which is important for the BN prototype. For the CH prototype, higher values of $\mathcal{S}_1^C$ correspond to balance between higher $\mathcal{G}_m$ for elements at the rightmost portion of the periodic table and lower values of $\epsilon_{\rm soc}^{\rm lu}$ for the \textit{p}-block elements composing the materials. In the $\rm WTe_2$ prototype, the first term of $\mathcal{S}_2^C$ separates the Janus-type materials inside the prototype. At the same time, both descriptors of model C are needed for classifying $\rm MoX_2$ (X = S, Se, Te) and $\rm WX_2$ (X = Se, Te) materials.

For GaS materials, the $\mathcal{S}_2^C$ descriptors balance the two terms. For topological oxides such as $\rm M_2O_2$ (M = Ir, Rh, W), the second term is predominant since the first term is small due to small $\epsilon_{m}^{\rm ho}$ values. In contrast, for $\rm Au_2Te_2$, the first term dominates, the ionization potential is very similar between the two atoms, decreasing $\tilde{{\rm IP}}_\sigma/\chi_M$ and $\mathcal{S}_2^C$.

In a simple model, one could use only these descriptors for making predictions. For this, an accurate classifier is needed, i.e., a rule (in this case, a line) that can distinguish between the two classes. This can be easily achieved by using a Support Vector Classifier (SVC) \cite{Vapnik1995} in the maps for each prototype and evaluated by cross-validation (CV). The SVC algorithm finds the best hyperplane to divide the two classes \cite{Schleder-MLreview}. The training uses only the expressions $\mathcal{S}_1$ and $\mathcal{S}_2$  as features for each data point. This can only be performed for the three prototypes \ch{CH}, \ch{GaSe} and \ch{WTe2} because they are the only ones with enough data to form a convex hull of the two classes. Training an SVC for the other structural prototypes may result in non-accurate generalization of the model. 
%
As seen in Fig. \ref{fig:sisso_convex_hull}, using the SISSO obtained maps has limited performance, given the simplicity of using only two descriptors in the models obtained. For instance, the SISSO descriptors are sufficient for separating the topological class in models A and B. However, in model C, the FeOCl prototype cannot be properly classified, see Fig. \ref{fig:sisso_convex_hull}c.
Nonetheless, the found descriptors provide much information in their non-linear combination of features, which can be further explored when combined with algorithms that enable additional complexity to be exploited.

\subsection{Machine learning classification using XGBoost tree method}

To increase the generalization capability of our model, we follow an approach similar to ref. \cite{gabriel_stability}. We include into our primary feature space the 10 top-ranked 2D descriptors found by SISSO that result in the largest absolute distance between the classes (in the case of the non-overlapping convex hull, models A and B) or the smallest overlap between classes (in the case of overlapping convex hull, model C). This results in 16 new features for model A, 13 new features for model B, and 8 new features for model C. 
Also, we include the prototype feature by a simple label encoding since this is our only categorical feature. 
We proceed to use this new data set with an extreme gradient boosting (XGBoost) tree classifier \cite{Chen2016_xgboost} via the scikit-learn wrapper \cite{scikit-learn}. This method performs the subsequent application of a tree classifier to correct the wrong predictions of the previous tree by introducing penalization for wrongly predicted labels. The extreme part comes from a series of efficient implementations that increase the accuracy and scalability of the method, and  most importantly, it implements regularization parameters and pruning of the trees, decreasing over-fitting.
We also tested the categorical structural prototype features by one-hot-encoding, which creates a binary column for each category, thus returning a sparse matrix of features. However, combined with the XGBoost algorithm, this led to decreased performance since the tree splitting is also sparse because it relies on few features.

\begin{figure*}[!htb]
    \centering
    \includegraphics[width=\linewidth]{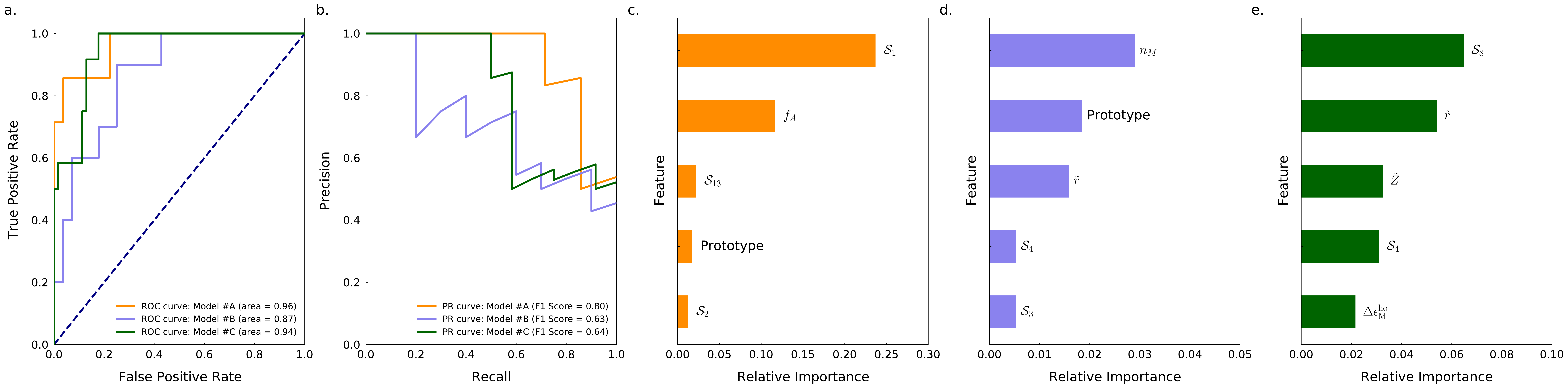}
    \caption{XGBoost models performance evaluation. (a) Receiver operating characteristic (ROC) curve with the corresponding area under curve (AUC) score. TPR is the true positive rate (also called recall or sensitivity), given by $TP / (TP+FN)$ and FPR is the false positive rate (also called specificity), given by $FP / (FP+TN)$. Values closer to 1 indicate better performance. (b) Precision-recall curve, given by $TP / (TP+FP)$ and $TPR$ respectively, with the corresponding F1 score, defined as their weighted average. (c-e) Relative feature permutation importance of the five most important features, for models A, B, and C.  }
    \label{fig:xgboost_roc_importance}
\end{figure*}

We split our data set into training and test in a stratified shuffled manner with a proportion of 80/20, which keeps the class balance between training and test. Then, the hyperparameter optimization was performed with grid search and 5-fold CV using the training set. The main final parameters for model A are a learning rate of 0.06, 27 estimators, maximum tree depth of 19, 0.85 column subsampling, and the $\ell_1$ and $\ell_2$-regularization parameters as 0.14 and 0.41, respectively. For the other models, the parameters and further details can be found in the Supplementary Material. 
We evaluate the models' performance using the area under the curve (AUC) of the receiver operating characteristic (ROC) curve, see. Fig. \ref{fig:xgboost_roc_importance}a. The ROC curve analyses the models' true positive rate against false positive rate with changing classification threshold. A model with perfect classification results in an AUC score equal to 1, and a model with random classification (dashed line) equal to 0.5. We obtain an AUC score of 0.96 for model A, 0.87 for model B, and 0.94 for model C, showing good performance for unseen data. 
In Fig. \ref{fig:xgboost_roc_importance}b we present the precision-recall (PR) curve for the three models. The PR curve measures the trade-off between precision (accurate results) and recall (target positive results). For higher recall values, the precision is higher for model A than for models B and C. As expected from the curves, the F1 score of 0.80 for model A is higher than for models B and C, of 0.63 and 0.64, respectively.
This indicates the trade-off between prediction accuracy and generalization capacity since models B and C have a broader prediction scope of their materials spaces.

To evaluate what features most contribute to predicting the target property, we obtain the permutation feature importance in Fig. \ref{fig:xgboost_roc_importance}c,d, and e for models A, B, and C, respectively.
We note that SISSO descriptors are crucial for classification. Indeed, for model A, the XGBoost model without SISSO descriptors reaches an AUC score of only 0.74, while models B and C without SISSO achieve decreased scores of 0.85 and 0.82. 
The structural prototype indeed plays a significant role in topology, as seen by the 4$^{\rm th}$ most important feature in model A and the 2$^{\rm nd}$ in model B. This is expected since the single-task SISSO method---without separating structural prototypes into different tasks---cannot perform the classification. For model A, given the convex hull in Fig. \ref{fig:sisso_convex_hull}a, it is expected that $\mathcal{S}_1$ is of great importance since it can almost separate both classes by itself.
Besides the SISSO descriptors already presented, the descriptor $\mathcal{S}_{13}^A$ in Eq. \eqref{eq:A:sisso_desc13}, is important for the classification of model A. Both $\mathcal{S}_2^A$ and $\mathcal{S}_{13}^A$ contain information about the difference in eigenvalue energy with and without SOC interaction for the A atoms. 
The other descriptors included in the models are presented in the Supplementary Material. 
For model B, the structural prototype is even more important in defining the topological character, reflecting the main challenge of the ML models in treating different structures while being able to generalize. 
For model C, the first SISSO features are not able to classify the topological systems for different prototypes, as seen in Fig. \ref{fig:sisso_convex_hull}c.
Correspondingly, other features are among the most important for the model.
Both for models B and C, the atomic properties are more important, of which valence and atomic radius are highlighted, while $\mathcal{S}_{3}^B$, $\mathcal{S}_{4}^B$, $\mathcal{S}_{4}^C$, and $\mathcal{S}_{8}^C$ in Eqs. \eqref{eq:B:sisso_desc3} to \eqref{eq:C:sisso_desc8} additionally take energies, element fractions, and periodic group into consideration.
Remarkably, XGBoost relies on a small descriptor space, which again shows the importance of adding information-rich descriptors such as those found by SISSO, allowing for the simple physical reasoning above. 

\begin{align}
    \mathcal{S}_{13}^A &= \frac{\epsilon^{ho}_A-\epsilon^{lu}_A}{|\epsilon^{lu}_{A}-\epsilon^{lu}_{soc\;A}|} \label{eq:A:sisso_desc13} \\
    \mathcal{S}_3^B &= {\rm IP}_M \times \tilde{r} \times \epsilon^{\rm ho}_{{\rm soc} \; M}/\mathcal{G}_m \label{eq:B:sisso_desc3}\\
    \mathcal{S}_4^B &= \bar{{\rm IP}}^2 \times \tilde{r}/\mathcal{G}_m \label{eq:B:sisso_desc4}\\
    \mathcal{S}_4^C &= \mathcal{G}_m\times\bar{\epsilon}^{\rm lu}_{\rm soc}/(f_M \times \epsilon^{\rm lu}_{{\rm soc},m}) \label{eq:C:sisso_desc4}\\
    \mathcal{S}_8^C &= |\tilde{{\rm IP}}\times[n]-{\rm EA}_M \times \bar{\mathcal{G}}| \label{eq:C:sisso_desc8}
\end{align}

\subsection{Prediction of new QSHIs and screening for selected candidates}

\subsubsection{Small-scale predictions: heuristic new combinations from the periodic table}

Having a machine learning model trained and validated, now we want to use our model to predict new QSHIs. Our initial approach is to make new combinations using the periodic table respecting simple heuristics of charge neutrality and structural prototype stoichiometry. We constraint our predictions to the \ch{CH}, \ch{GaSe} and \ch{WTe2} prototypes, finding 569 new topological materials for model A. To improve the computational efficacy for calculating the properties of these candidates, we make use of the thermodynamic stability model of ref. \cite{gabriel_stability}. We screen for highly stable materials, i.e., those that have formation energies ($\Delta H_{f}$) and energy above the convex hull ($\Delta H_{hull}$) below \SI{0}{\electronvolt} and \SI{0.2}{\electronvolt}, respectively, and materials with medium stability, with $\Delta H_{f} < \SI{0}{\electronvolt}$ and $\Delta H_{hull} > \SI{0.2}{\electronvolt}$, which could possibly be stabilized depending on the synthesis route and substrate.
The thermodynamic stability model proved to be a powerful and reliable tool for screening our predictions. The screening resulted in 138 candidates.

At this stage, one should note one specificity of our approach. We initially choose to train our model for the topological classification of insulators. Although this is a reasonable choice for training and decreasing the complexity of our initial model, this turns out to be difficult for predicting unknown materials since we do not know if combining elements in a particular prototype will result in an insulator or metal. We also lack a proper structure guess for the screened candidates.

We performed the DFT calculations for the 138 candidate compounds using an initial structural guess of a representative candidate from the corresponding prototype. The parameters used in the DFT calculations of the screened candidates are found in the Methods section. 

\begin{figure*}[!ht]
    \centering
    \includegraphics[width=\linewidth]{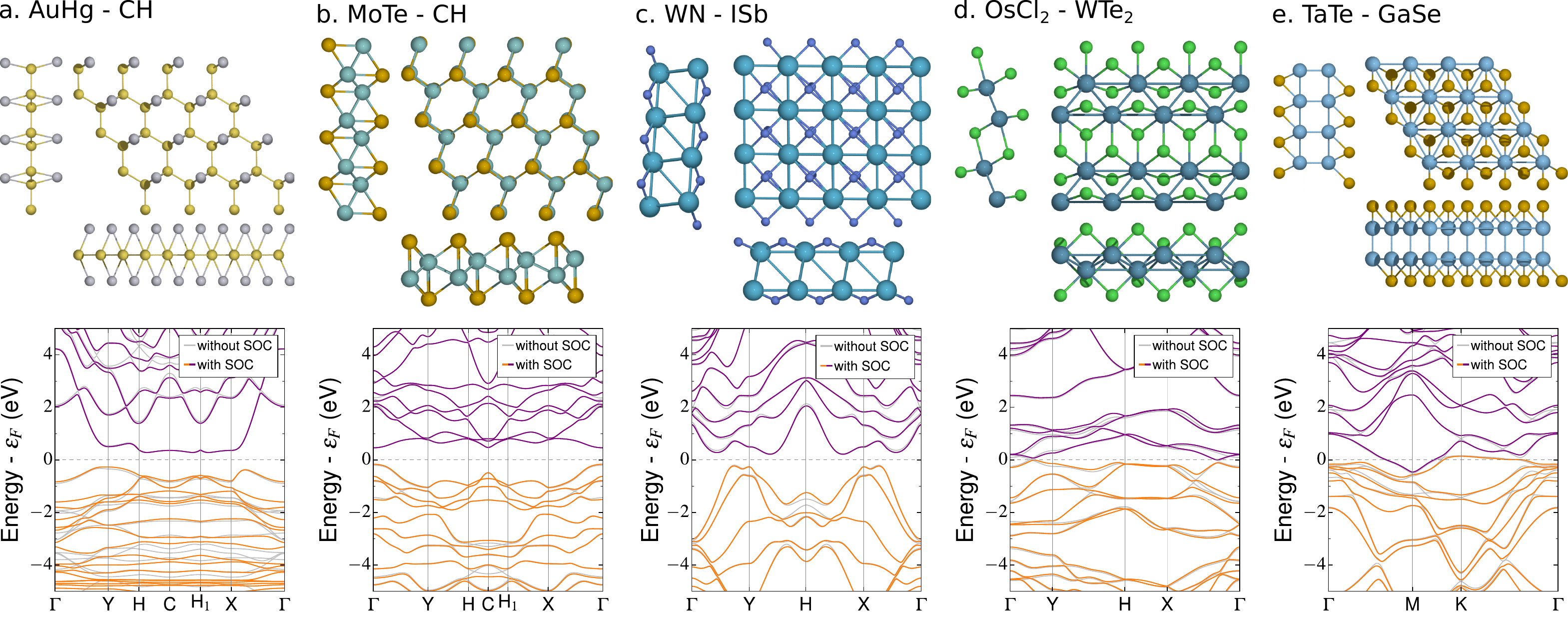}
    \caption{Representative electronic band structure from calculated candidates. Large gap TIs: (a) AuHg in the CH prototype, (b) MoTe in the CH prototype, and (c) WN with ISb as the most similar prototype. (d) \ch{OsCl2} is a TI with a small band gap in the \ch{WTe2} prototype, and (e) TaTe is a direct gap metal with $\mathbb{Z}_2 = 1$ in the \ch{GaSe} prototype. For insulators (a)--(d), the different colors correspond to occupied and unoccupied bands. In the case of DGM (e), the different colors correspond to the bands considered for invariant calculation. The given prototype corresponds to the initial structure before structural relaxation.}
    \label{fig:bands_examples}
\end{figure*}

We obtain 109 metallic compounds and 29 insulators; from these, 11 are topological insulators. These materials are summarized in Table \ref{tab:predictions}. 
We found 9 TIs unreported in the literature to the best of our knowledge. 
In ref. \cite{Wang2017a}, \ch{AsO} is reported as a QSHI highly robust to strain and compatible with hBN substrate. Ref. \cite{Xiaofeng2014a} reports several TMDCs in the 1T$^\prime$ phase. Many of them are topological, including WS$_2$ as reported here. Also, WS$_2$ can be found on C2DB, but is classified as a metal even though the PBE-SOC band structure shows an energy gap.
We find 2 materials with a band gap larger than \SI{0.5}{\electronvolt}. 
In Fig. \ref{fig:bands_examples}(b) we demonstrate the band structure of \ch{MoTe} and in Fig. \ref{fig:bands_examples}(d) the band structure of \ch{OsCl2}, these are two representative cases in which the SOC contributes differently to the band structure. In \ch{MoTe}, the band gap is large, so we expect that the non-trivial topology arises as a band inversion induced by the SOC altering the band character. Conversely, in \ch{OsCl2}, the SOC opens a small inverted band gap, which is otherwise closed.

Investigating the 109 metallic materials, we found 85 candidates defined as direct gap metals (DGMs) \cite{marzari_new_TI}. In these materials, a metallic character arises from the Fermi level crossing occupied bands. Still, we can observe direct transitions between valence-like and conduction-like bands throughout the Brillouin zone. We only consider reasonable energy gaps larger than \SI{5}{\milli\electronvolt}. These direct gaps appear due to several reasons, including SOC interaction. As we observe for the representative candidate \ch{TaTe} in Fig. \ref{fig:bands_examples}(e), a direct gap separates the ``valence'' and ``conduction'' bands near the Fermi level only when SOC is included. For DGMs, the $\mathbb{Z}_2$ invariant is well defined as long as there are no band crossings throughout the BZ and can be computed for the selected number of lower energy bands. From these 85 candidates, we obtain 39 topological DGMs (TDGMs) which feature $\mathbb{Z}_2 = 1$. The predictions thus result in a total of 50 topological materials. See the Supplementary Material for relaxed geometries and band structure of these compounds. Although some of these candidates can be found on C2DB with a similar structural prototype, they possess different atomic and electronic structures, ensuring their novelty and substantially different properties.

\begin{table}[h!]
    \centering
    \caption{Model A predicted $\mathbb{Z}_2 = 1$ insulators confirmed by DFT calculations from the new combinations generated. The given prototype corresponds to the initial structure before structural relaxation. The corresponding Kohn-Sham electronic band gap at the PBE level of theory is also presented.}
    \label{tab:predictions}
    \begin{tabular}{lcc}
    \hline
    \multicolumn{1}{c}{Material} & \multicolumn{1}{c}{Prototype} & \multicolumn{1}{c}{KS gap (meV)} \\ \hline
    MoTe &        CH &   625 \\
    AuHg &        CH &   549 \\
     AsO &        CH &   88 \\
    TeAg &        CH &   85 \\
    MoSi &        CH &   41 \\
    TiSe &        CH &   12 \\
    TiTe &        CH &   9 \\
    BaCl &      GaSe &   66 \\
  WS$_2$ &   WTe$_2$ &   26 \\
 BaO$_2$ &   WTe$_2$ &   13 \\
OsCl$_2$ &   WTe$_2$ &   9 \\
        \hline
    \end{tabular}
\end{table}


\begin{figure}[!htb]
    \centering
    \includegraphics[width=\linewidth]{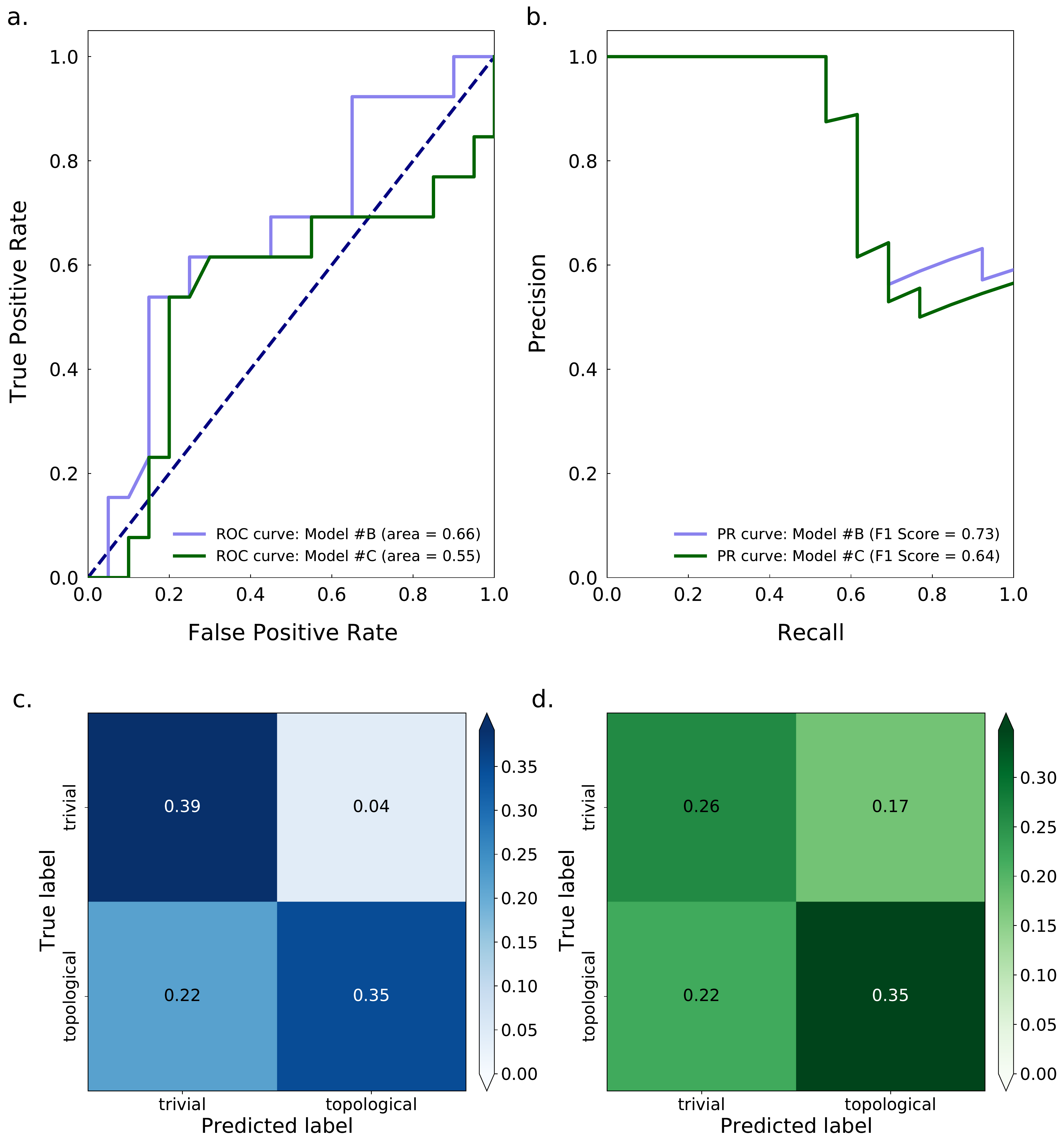}
    \caption{Validation of models B and C on the previously generated and calculated materials. In (a) we present the receiver operating characteristic (ROC) curve with the corresponding area under curve (AUC) score. TPR is the true positive rate and FPR is the false positive rate. (b) Precision-recall curve with the corresponding F1 score. (c) and (d) present the confusion matrix for models B and C, respectively.}
    \label{fig:predict_calculated}
\end{figure}

In Fig. \ref{fig:predict_calculated} we present the validation of models B and C predictions on the generated and calculated materials. The AUC score from the ROC curves for both models shows only average performances of 0.66 and 0.55, respectively. Furthermore, the F1 scores of 0.73 and 0.64 also show average performance, showing that in order to obtain larger recall values (i.e., retrieving all relevant materials for the property), it is expected that the precision of the models will be lower, i.e., they will also predict a significant fraction of false positives.
Indeed, in the confusion matrix for each model, Figs. \ref{fig:predict_calculated}c,d, we see that most of the predictions are correct, but with a fraction of false negatives (the majority for model B) and false positives.
Therefore, we can once again see that more broad models that include more prototypes are a trade-off between prediction and generalization.

\subsubsection{Larger-scale predictions: systematic databases}


An important validation of our model's generalization is the prediction on other databases, where we can take advantage of already known structures and band gaps. The remaining obstacle is the prototype classification which will be addressed in the following.

First, using model A, we performed the predictions on the 2DMatPedia database \cite{Zhou2019}, which has 6351 entries. We screened for insulators (3599), non-magnetic (2646), binary (1606) materials. We then excluded repeated entries similar to those on the training/test set by comparing chemical formula, composition, and stoichiometry, resulting in 822 candidates. Finally, we restricted our candidates to 6 atoms in the unit cell, consistently with the training prototypes for topological classification. Our final prediction set contains 164 materials.

The next step is to assign a prototype for each entry. For this, instead of manually inspecting each structure, we generated structural features using the crystal nearest-neighbor (CNN) fingerprint method, which uses information of the NN local environment for generating a fingerprint for each structure site; these site features are used to create structure features by statistics, as mean, standard deviation, minimum and maximum of site features \cite{Zimmermann2020,Zimmermann2017,Ward2018}. Finally, we add to these features the total number of atoms and the number of different chemical species in the formula.

Using the structural features obtained by the CNN fingerprint method, we can measure the similarity between each structure and representative candidates for each prototype, assigning the most similar prototype, also allowing for an unknown prototype in the case of a distance greater than a threshold. We use the cosine similarity measure. The structure is considered similar to one of the prototypes if the metric is at least 0.9. The prototype is assigned based on the greatest similarity between the structure and the material used as prototype representation. This is performed directly on our screened 164 materials.

We predict that 115 out of 164 materials are presented in the prototypes used for training the topological classification model. From these 115, we restrict our search for materials with an electronic band gap smaller than 0.8 eV. 
We can now predict the topological class of these materials using our three models.
Using model A, we predict 12 topological materials, including buckled antimonene (Sb), which is not on the C2DB and is topological for lattice constant greater than \SI{4.7}{\angstrom} \cite{Zhu2019}. From these 11 predicted unknown materials, after calculations, we verify 5 topological materials. Employing models B and C, 7 and 4 materials are predicted as topological, and we verify 2 as topological insulators for each.
Table \ref{tab:predictions_2dmatpedia} summarizes our predictions for the 2DMatPedia database, and the predicted large gap material WN is presented in Fig. \ref{fig:bands_examples}c. One should also note that the attributed prototypes are based on structure similarity since the known used prototypes for 2D materials are limited, one should not expect full compatibility with the actual prototypes representatives, but the structural similarity is enough for ensuring the application of the topological classification method, also putting to test the generalization of our model. The geometries and band structures can be found on the Supplemental Material. In some cases, the final structure after relaxation belongs to a different crystal system of the initial prototype, which also occurs for the C2DB database.
We here also mention that we employed our process to the V2DB \cite{Sorkun2020}, a database containing 316,505 generated 2D materials covering a large compositional space, with predicted physical properties. 
The 15 new materials predicted as topological resulted in metallic systems from the DFT calculations from our screening described above. 
Therefore, we verify that the database predicted band gap does not agree with the calculations for our studied materials.

\begin{table}[!ht]
    \centering
    \caption{Models predicted $\mathbb{Z}_2 = 1$ insulators confirmed by DFT calculations from the 2DMatPedia database \cite{Zhou2019}. The given prototype corresponds to the most similar one as calculated by the crystal nearest-neighbor (CNN) fingerprint. The corresponding Kohn-Sham electronic band gap at the PBE level of theory is also presented.}
    \label{tab:predictions_2dmatpedia}
    \begin{tabular}{lp{1.3cm}p{1.3cm}cc}
    \hline
        \multicolumn{1}{c}{Material} & Most similar prototype  & Predicted by model & $E_g$ (meV) & 2DMatPedia ID \\ \hline
WN	&	ISb	&	A	&	453	&	\href{http://www.2dmatpedia.org/2dmaterials/doc/2dm-2568}{2dm-2568}	\\
MoAs	&	ISb, CH*	&	A, B	&	203	&	\href{http://www.2dmatpedia.org/2dmaterials/doc/2dm-2128}{2dm-2128}	\\
\ch{SbBr5}	&	ISb, CH	&	A, B, C	&	80	&	\href{http://www.2dmatpedia.org/2dmaterials/doc/2dm-2745}{2dm-2745}	\\
\ch{SbCl5}	&	CH	&	C	&	19	&	\href{http://www.2dmatpedia.org/2dmaterials/doc/2dm-2214}{2dm-2214}	\\
HfBr	&	WTe$_2$	&	A	&	128	&	\href{http://www.2dmatpedia.org/2dmaterials/doc/2dm-662}{2dm-662}	\\
\ch{WSb2}	&	WTe$_2$	&	A	&	125	&	\href{http://www.2dmatpedia.org/2dmaterials/doc/2dm-2905}{2dm-2905}	\\

\hline
    \end{tabular}
\end{table}

Since the accuracy of our models is limited, the number of predicted topological materials could be increased by choosing a less conservative baseline for making predictions and increasing the number of materials predicted in the positive class, that is, increasing the recall while decreasing the precision.  However, this is a decision taken when creating the ML model. In our case, we chose the ROC AUC as the hyperparameter optimization metric as we strive for a balance between precision and recall. If type II errors (false negative or ``miss'') are to be avoided, this can be favored by choosing an appropriate metric but at a tradeoff with resource consumption and efficiency.

As demonstrated in \cite{Vidal2011}, for 3D topological insulators, false-positives of topological phases identified with DFT may result from the strong underestimation of the energy gap or incorrect choice of the ground-state structure, sensitive to the lattice parameters optimized with a given exchange and correlation functional \cite{Vidal2011,Malyi2020a}. As such, our predictions of Tables \ref{tab:predictions} and \ref{tab:predictions_2dmatpedia} correspond to the GGA-PBE DFT functional, and can vary for different levels of theory.

\section{Conclusions and outlook}

In the present work, our goal is to use data-driven approaches for the accelerated discovery of topological phases of matter. We focus on topological insulators and using the available databases and literature. We employ an automated feature engineering method that produces simple descriptors allowing physical reasoning for the classification task. Using the generated descriptors, we apply a machine learning technique to predict the topological classification of insulating materials. Combining both methods allows the XGBoost model to perform accurate predictions using a restricted descriptor space that allows for physical reasoning.

We first make predictions for new combinations of elements from known structural prototypes. This turns out to be challenging since most reasonable combinations of elements are already known. To reduce the number of candidates, we apply the ML method of ref. \cite{gabriel_stability} to screen for medium and highly stable 2D materials. 
We make a second prediction round, now using the 2DMatPedia database, finding 6 new topological materials, and testing the generalization of our model. 
In total, we successfully identify 56 topological materials, of which 17 are quantum spin Hall insulators, and 9 are unreported in the literature, with 3 presenting a large energy gap suitable for room-temperature applications.
Considering high-quality topological insulators having electronic band gaps above 25 meV, we reach 12 materials, an encouraging number considering the many years taken to achieve the approximately 20 high-quality discovered 2D topological insulators to date \cite{Vishwanath2019_2Dprb}.

We show that the combined use of SISSO and extreme gradient boosting methods allow for efficient ML algorithms with high interpretability and generalization capability. These are desirable qualities for data-driven methods applicable to materials science.
We also present a detailed analysis of the importance of verifying each choice regarding creating machine learning models, which will affect their results, applicability, and generalization. 
Of the different variables explored, we consider the effect of the initial dataset, the screening choices, materials stability, representations, ML algorithms, and finally, the transferability between models and databases. 

Considering that historically discoveries are in many instances guided by intuition rather than being entirely rational, artificial intelligence and machine learning, in particular, are powerful methods to aid in automatically extracting insights from past collected data and making significant (and sometimes non-obvious to human counterparts) advances in the future.

\section*{Methods}\label{section:methods}

\subsection*{Density functional theory calculations}

All density functional theory (DFT)\cite{dft1964,dft1965,Schleder-MLreview} calculations were performed with the aid of the ASE package \cite{ase-paper,early-ase-paper} using the Vienna \textit{ab initio} Simulation Package (VASP) \cite{vasp1,vasp2}. We use the generalized gradient approximation (GGA) \cite{gga} with Perdew-Burke-Ernzerhof (PBE) \cite{pbe} exchange and correlation functional and the projector augmented-wave (PAW)\cite{paw} method for describing ionic potentials. All calculations are performed without and with self-consistent spin-orbit coupling (SOC). We used \SI{600}{\electronvolt} for the plane-waves kinetic energy cutoff.

\subsection*{Atomic energy features}

To compute the atomic eigenvalues we use a single atom inside an asymmetric box. The atom is slightly displaced from the box center and enough vacuum space, at least \SI{13}{\angstrom}, is included in all directions to avoid interactions with periodic images. To achieve convergence and integer occupation we enforce the correct occupation when needed. All eigenvalues and total energies are shifted to the vacuum reference level.

\subsection*{New materials}

All 2D materials are computed using \SI{20}{\angstrom} of vacuum space to avoid interactions with periodic images. We first perform the relaxation until reaching a force criterion of \SI{e-2}{\electronvolt\per\angstrom}. We follow by calculating the ground-state (GS) charge density and then the band structure, both without and with SOC interaction. The relaxations were performed with k-point density of $6\;/$\si{\per\angstrom} and GS charge density with $12\;/$\si{\per\angstrom}. The band structures were computed using 201 points along the high-symmetry path.

\subsection*{$\mathbb{Z}_2$ invariant calculation}

The $\mathbb{Z}_2$ invariant is computed tracking the evolution of the Wannier charge centers (WCCs) \cite{z2pack_1,z2pack_2} as implemented in the Z2Pack code \cite{z2pack_code}. 

\section*{Data Availability}

The data that supports the findings of this study are available within the article and its supplementary material. 
The equations of the descriptors found by SISSO and the XGBoost topological classification model parameters are given in the Supplementary Material (SM). 
We make available our final XGBoost classification model as a python pickle file. 
Also in the SM, we make available all structures images and corresponding band structure for the predicted topological materials. In a zip file we include structure files (as POSCAR), band structures in ASE json format and a csv with the atomic features calculated with DFT, the other features are acquired using the mendeleev package \cite{mendeleev2014}.
Additional data are available from the corresponding author upon request.


\begin{acknowledgments}
The authors acknowledge insightful feedback from Prof. Marcio Costa and Prof. Efthimios Kaxiras.
This work is partially supported by São Paulo Research Foundation (FAPESP), grants no. 2017/18139-6, 2019/04527-0, and 2017/02317-2, and Brazilian funding agency CNPq. The authors acknowledge the SDumont high-performance computer at the Brazilian National Scientific Computing Laboratory (LNCC) for computational resources.
\end{acknowledgments} 

\section*{AUTHOR CONTRIBUTIONS}

All authors conceived the project.
B.F. performed the ML models and DFT simulations with G.R.S. feedback.
All authors analyzed and discussed the results.
G.R.S. and B.F. wrote the initial manuscript, which all authors revised.

\section*{COMPETING INTERESTS}
The authors declare no competing interests. 

\bibliographystyle{apsrev4-2}
\bibliography{references}
\end{document}